\documentclass[10pt,letterpaper,twocolumn]{article} 

\usepackage{ol2}
\usepackage[draft]{hyperref}
\usepackage{amsmath}

\usepackage{graphics}

\begin{document}

\twocolumn[ 

\title{Absolute detector calibration using twin beams}


\author{Jan {Pe\v{r}ina,~Jr.},$^{1,*}$ Ond\v{r}ej Haderka,$^2$
V\'{a}clav Mich\'{a}lek,$^2$ and Martin Hamar$^{2}$}

\address{
$^1$ RCPTM, Joint Laboratory of Optics of Palack\'{y} University
and Institute of Physics of Academy of \\ Sciences of the Czech
Republic, Faculty of Science, Palack\'{y} University, 17.
listopadu 12, \\ 77146 Olomouc, Czech Republic
\\
$^2$ Institute of Physics of Academy of Sciences of the Czech
Republic, Joint Laboratory of Optics of \\ Palack\'{y} University
and Institute of Physics of Academy of Sciences of the Czech
Republic, \\ 17. listopadu 12, 772 07 Olomouc, Czech Republic
\\
$^*$Corresponding author: perinaj@prfnw.upol.cz}

\begin{abstract}
A method for the determination of absolute quantum detection
efficiency is suggested based on the measurement of photocount
statistics of twin beams. The measured histograms of joint
signal-idler photocount statistics allow to eliminate an
additional noise superimposed on an ideal calibration field
composed of only photon pairs. This makes the method superior
above other approaches presently used. Twin beams are described
using a paired variant of quantum superposition of signal and
noise.
\end{abstract}


 ] 

The first suggestions to use photon pairs in absolute detector
calibration have occurred soon after the experimental evidence of
emission of photons in pairs in the process of spontaneous
parametric down-conversion had been given \cite{Burnham1970}.
The suggested method is based on the fact that both photons are
created in the nonlinear process
simultaneously. 
Provided that one
(signal) photon is detected at its (signal) detector,
we know for sure in the ideal case that the second photon exists in the beam \cite{Malygin1981}. Thus it impinges on a
(idler) detector that is calibrated with probability one. Many
repetitions of the experiment then provide the required absolute
quantum detection efficiency (QDE) \cite{Migdall1999}. Following
this simple scheme, absolute QDE $ \eta_i $ of an idler detector
is reached along the formula $ \eta_i = \langle m_s m_i \rangle /
\langle m_s \rangle $, where $ \langle m_s m_i \rangle $ gives the
number of coincidence counts at both detectors and
$ \langle m_s \rangle $ determines the number of signal-detector
single counts. Weak photon fields having only a small fraction of
a photon in a detection window on average are needed in this
approach. The method automatically gives also QDE of the signal
detector. It can be used for the calibration of both analog and
photon-counting detectors with the precision comparable to other
metrology methods exploiting thermal and semiconductor detectors
\cite{Brida2006}.

The presented method, however, requires as 'the probe' a weak
field composed only of photon pairs. It is unable to cope with any
additional noise in the form of un-paired single photons
superimposed on the paired field. A simple analysis shows that the
presence of additional single-photon noise counts $ \langle m_i
\rangle_{\rm noise} $ modifies the formula for QDE $ \eta_i $; $
\eta_i = \langle m_s m_i \rangle / \langle m_s \rangle - \langle
m_i \rangle_{\rm noise} $. However, if single-photon noise is
present in both fields, we cannot partition it from the paired
part of the probe field. The improved formula QDE $ \eta_i $ is
thus not useful. Single-photon noise can only be eliminated as
much as possible using a careful experimental arrangement.

Despite this drawback, the method has been generalized to include
also stronger photon-pair fields \cite{Brida2010} in which both
pump-field intensity fluctuations and transverse correlations of
photons in a pair play an important role.
If the ratio of signal and idler QDEs is known and only a paired
field is assumed, a simple relation between QDE and
noise-reduction factor (quantifying sub-shot-noise photon
correlations) can be revealed \cite{Brida2010}.

As we show in this letter, additional single-photon noise in the
probe field can be eliminated considering photon-number resolving
detectors like intensified CCD (iCCD) cameras
\cite{Haderka2005a,Hamar2010}, array detectors \cite{Afek2009} or
electron-multiplied CCD (EMCCD) cameras \cite{Zhang2009}.
This then allows to determine both signal and idler QDEs with, in
principle, no precision limitations. The method is based on the
measurement of joint signal-idler photocount distribution (JPCD)
\cite{Perina2007}. In detail, the experimental JPCD is fitted
assuming a special form of the probe field derived in the
framework of superposition of signal and noise applied to paired
fields \cite{Perina2005,Perina2006}. This fitting uses both first
and second photocount moments and minimizes declinations from the
experimental photocount histogram \cite{Worsley2009}. This
approach provides signal and idler QDEs, together with parameters
of the noisy single-photon signal and idler fields giving a
detailed detection characterization \cite{DAuria2011}.

We demonstrate the general method by considering the measurement
performed by an iCCD camera Andor. Multi-mode twin beams at the
wavelength around 560~nm filtered by spectrally rectangular 14-nm
wide (FWHM) interference filter have been generated by the pulsed
third harmonics of a Ti:sapphire laser tuned at 840~nm in a
non-collinear type-I interaction in a 5-mm long BaB$ {}_2 $O$ {}_4
$ crystal (for details, see \cite{Hamar2010}). Whereas the signal
field has been directly sent to the photocathode of the camera,
the idler field has been reflected on a dielectric mirror
(reflectivity 99.2\%) first and then impinged on a different area
of the photocathode. This causes asymmetry in the signal ($ \eta_s
$) and idler ($ \eta_i $) QDEs. Using pulsed pumping and repeating
the measurement $ \sim 10^5 $ times, we have arrived at histogram
$ f(c_s,c_i) $ giving the number of measurements with the
registered $ c_s $ signal and $ c_i $ idler photocounts. In the
experiment, also the level of dark counts has to be monitored in
order to allow for subtracting this additional, but independently
quantified, noise.

A suitable choice of a general six-parameter form of the joint
signal-idler photon-number distribution (JPND) together with
appropriate values of signal ($ \eta_s $) and idler ($ \eta_i $)
QDEs lies in the heart of the method. However, according to the
theory of measurement, only values of the first and second moments
of measured quantities are reliably determined after a reasonable
number of measurement repetitions. Including the five measured
values of the first and second moments, three free parameters
remain in the considered eight-parameter problem. Our
investigations have shown that the values of remaining three
parameters are successfully revealed requiring the best fitting of
the theoretical JPND to the experimental JPCD.

In the first step, we determine the first ($ \langle c_s \rangle
$, $ \langle c_i \rangle $) and second ($ \langle c_s^2 \rangle $,
$ \langle c_s c_i \rangle $, $ \langle c_i^2 \rangle $) moments of
the measured numbers of photocounts. Knowing the first ($ \langle
d \rangle $) and second ($ \langle d^2 \rangle $) moments of the
dark-count distribution, its contribution to the measured
photocount moments can be eliminated and moments of detected
photoelectrons can be, step by step, found ($ a=s,i $):
\begin{eqnarray}   
 \langle m_a\rangle &=& \langle c_a\rangle - \langle d_a \rangle ,
   \nonumber \\
 \langle m_a^2\rangle &=& \langle c_a^2\rangle - 2 \langle m_a\rangle
  \langle d_a \rangle - \langle d_a^2 \rangle ,
\label{1}    \\
 \langle m_s m_i\rangle &=& \langle c_s c_i \rangle - \langle m_s\rangle
  \langle d_i \rangle - \langle m_i\rangle \langle d_s \rangle -
  \langle d_s d_i \rangle .  \nonumber
\end{eqnarray}

On the other hand, the probe photon field 'in front of a detector'
can be considered as composed of three independent parts: pairs,
signal noisy photons and idler noisy photons. They can be
characterized by their numbers $ M_p $, $ M_s $, and $ M_i $ of
equally-populated modes and mean photon numbers $ b_p $, $ b_s $,
and $ b_i $ per mode, respectively. The first and second
photon-number moments of these fields can be expressed as ($
a=p,s,i $):
\begin{eqnarray}    
 & & \langle n_a\rangle = M_a b_a, \hspace{5mm}
   \langle (\Delta n_a)^2 \rangle = M_a b_a (1+b_a).
\label{2}
\end{eqnarray}

QDEs $ \eta_s $ and $ \eta_i $ provide the bridge between the
'theoretical' photon-number moments and experimental photoelectron
moments. Quantum detection theory \cite{Perina1991} gives us ($
a=s,i $):
\begin{eqnarray}   
 \eta_a \left[ \langle n_p\rangle + \langle n_a\rangle \right] &=&
  \langle m_a\rangle , \nonumber \\
 \eta_a^2 \Biggl[ \langle (\Delta n_p)^2 \rangle + \langle (\Delta n_a)^2 \rangle
   + \frac{1-\eta_a}{\eta_a}\left( \langle n_p\rangle +
  \langle n_a\rangle \right) \Biggr] \hspace{-35mm} & & \label{3} \\
  &=& \langle (\Delta m_a)^2 \rangle , \nonumber \\
 \langle (\Delta n_p)^2 \rangle &=&
  {\langle \Delta m_s \Delta m_i \rangle }/({\eta_s\eta_i}).
\nonumber
\end{eqnarray}

If QDEs $ \eta_s $ and $ \eta_i $ were known, Eq.~(\ref{3}) would
give five constraints for the determination of six parameters $
M_a $ and $ b_a $. This represents a serious problem arising from
two points: (1) Only the first and second photoelectron moments
are experimentally available with sufficient precision and (2) The
probe photon field is non-classical due to its predominantly
paired character that enforces the use of at least six independent
parameters in its realistic description. The solution of
Eq.~(\ref{3}) can be expressed as a one-parameter system
conveniently parameterized by the mean photon-pair number $
\langle n_p\rangle $ ($ a=s,i $):
\begin{eqnarray}   
 \langle n_a\rangle &=& {\langle m_a\rangle}/{\eta_a} - \langle
   n_p\rangle , \nonumber \\
 \langle (\Delta n_a)^2 \rangle &=& \frac{ \langle (\Delta m_a)^2
  \rangle}{ \eta_a^2} - \frac{ \langle \Delta m_s  \Delta m_i
  \rangle}{ \eta_s\eta_i} -\frac{1-\eta_a}{\eta_a^2} \langle m_a \rangle ,
  \nonumber \\
 \langle (\Delta n_p)^2 \rangle &=& { \langle \Delta m_s  \Delta m_i
  \rangle}/( \eta_s\eta_i) .
\label{4}
\end{eqnarray}

However, also QDEs $ \eta_s $ and $ \eta_i $ are not known. As the
third and higher photoelectron moments cannot be reliably used for
the chosen number of measurement repetitions \cite{PerinaJr2012},
we suggest to minimize the declinations of experimental ($ f $)
and theoretical ($ p_c $) photocount distributions. A JPCD $ p_c $
can be derived from a JPND $ p $ provided that the detection
process is characterized. The JPND $ p $ of a field composed of
pair, signal, and idler components can naturally be written as a
convolution of three Mandel-Rice distributions \cite{Perina1991}:
\begin{eqnarray}  
 p(n_s,n_i) &=& \sum_{n=0}^{{\rm min}[n_s,n_i]} p(n_s-n;M_s,b_s)
   \nonumber \\
 & & \mbox{} \times p(n_i-n;M_i,b_i) p(n;M_p,b_p);
\label{6}
\end{eqnarray}
$ p(n;M,b) = \Gamma(n+M) / [n!\, \Gamma(M)] b^n/(1+b)^{n+M} $.
Following Eq.~(\ref{2}), numbers $ M_a $ of modes and mean
photon-numbers $ b_a $ per mode can be derived from the first and
second photon moments occurring in Eq.~(\ref{4}) as follows ($
a=p,s,i $):
\begin{eqnarray}   
 b_a = \frac{\langle (\Delta n_a)^2 \rangle }{ \langle n_a
  \rangle } -1 , \hspace{4mm}
 M_a = \frac{ \langle n_a \rangle^2 }{ \langle (\Delta n_a)^2\rangle -
  \langle n_a \rangle }.
\label{7}
\end{eqnarray}

A detailed theory appropriate for a detector with $ N $ pixels,
QDE $ \eta $ and dark-count rate $ D \equiv \langle d\rangle / N $
has been developed in \cite{PerinaJr2012} and provided the
probabilities $ T(c,n) $ of having $ c $ photocounts out of a
field with $ n $ photons:
\begin{eqnarray}     
  T_a(c,n) &=& \begin{pmatrix} N_a \cr c \end{pmatrix} (1-D_a)^{N_a}
  (1-\eta_a)^{n} (-1)^{c} \nonumber \\
   & & \hspace{-2.5cm} \times \sum_{l=0}^{c}
  \begin{pmatrix} c \cr l \end{pmatrix} \frac{(-1)^l}{(1-D_a)^l}
  \left( 1 + \frac{l}{N_a} \frac{\eta_a}{1-\eta_a}
   \right)^{n}, a=s,i .
\label{8}
\end{eqnarray}
Using the formulas in Eqs.~(\ref{6}) and (\ref{8}) the JPCD $ p_c
$ can be expressed as:
\begin{eqnarray}   
  p_c(c_s,c_i) = \sum_{n_s,n_i=0}^{\infty} T_s(c_s,n_s)
   T_i(c_i,n_i) p(n_s,n_i) .
\label{9}
\end{eqnarray}

Finally, the appropriate values of QDEs $ \eta_s $ and $ \eta_i $
and mean photon-pair number $ \langle n_p\rangle $ are obtained by
minimizing the function $ {\cal D} $ quantifying the declinations
of JPCD $ p_c $ and experimental histogram $ f $:
\begin{equation}  
 {\cal D} = \sqrt{ \sum_{c_s,c_i=0}^{\infty} \left[p_c(c_s,c_i) -
  f(c_s,c_i)\right]^2 }.
\label{10}
\end{equation}

\begin{figure}         
 \resizebox{0.495\hsize}{!}{\includegraphics{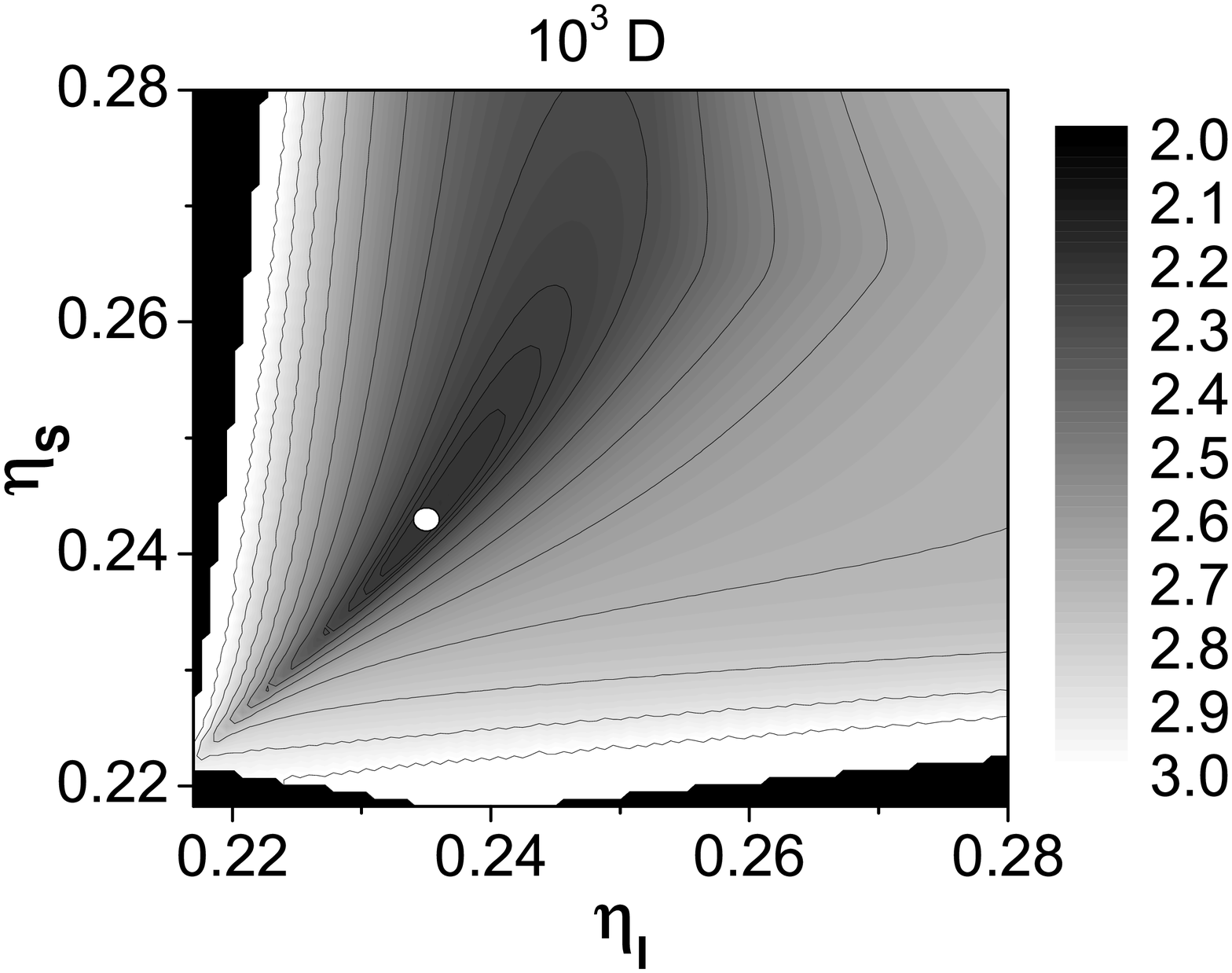}}
 \resizebox{0.495\hsize}{!}{\includegraphics{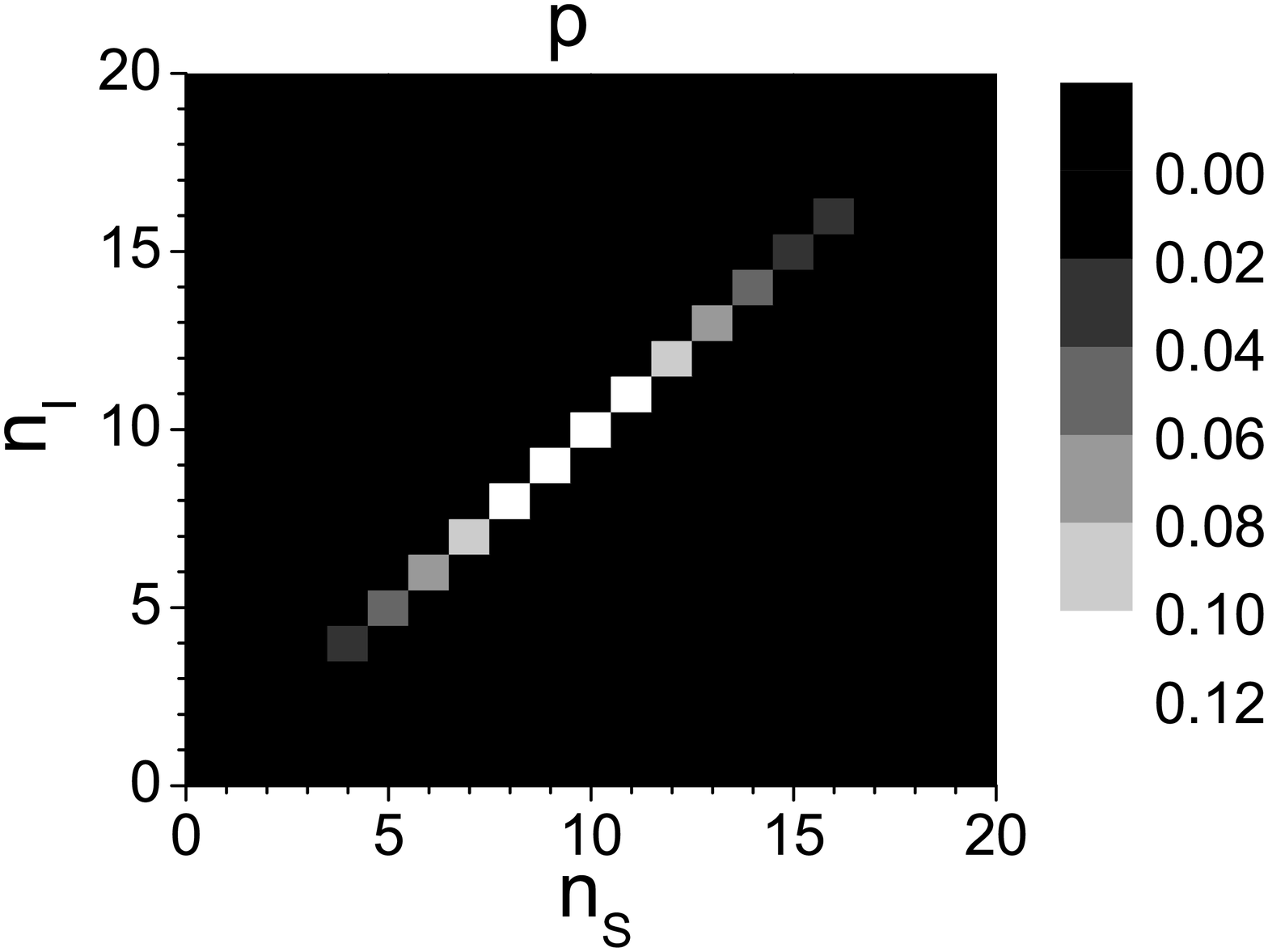}}
  \caption{Minimum of declinations $ {\cal D} $ determined
  over the allowed values of $ \langle n_p\rangle $ depending
  on QDEs $ \eta_s $ and $ \eta_i $. In black areas close to $ \eta_s $
  and $ \eta_i $ axes,
  Eqs. (\ref{3}) have no solution. For optimum values of QDEs
  $ \eta_s = 0.243 $ and $ \eta_i = 0.235 $, JPND $ p(n_s,n_i) $ is shown.}
\label{fig1}
\end{figure}

To practically demonstrate the power of the method, we discuss a
typical measurement described in Fig.~\ref{fig1}. Active area of
the used iCCD camera is composed of 128x128 independent pixels
with equal QDEs obtained by hardware binning of the original 1
megapixel resolution. Processing the raw data eliminates spatial
blurring of the detection spots. Signal and idler photons are
captured in different areas of the photocathode. Some pixels in
the photocathode are also reserved for monitoring the noise.
However, not all noises can be quantified this way. For example,
single photons arising from fluorescence in the nonlinear crystal
are difficult to monitor. There also occur pump-intensity
fluctuations that modify the photon-pair statistics. As an
example, we analyze the measurement that has resulted in the
following photoelectron moments after eliminating dark counts: $
\langle m_s\rangle = 2.411 \pm 0.002 $, $ \langle m_i\rangle =
2.353 \pm 0.004$, $ \langle (\Delta m_s)^2 \rangle = 2.489 \pm
0.003 $, $ \langle (\Delta m_i)^2 \rangle = 2.449 \pm 0.005 $, and
$ \langle \Delta m_s \Delta m_i \rangle = 0.597 \pm 0.003 $.
Applying the usual simplified method for the determination of QDE
[neglecting noises in (\ref{3}) and assuming $ \langle (\Delta
n_p)^2 \rangle = \langle n_p \rangle $], we arrive at the values $
\eta_s = \langle \Delta m_s \Delta m_i \rangle / \langle m_i
\rangle = 0.254 \pm 0.001 $ and $ \eta_i = \langle \Delta m_s
\Delta m_i \rangle / \langle m_s \rangle = 0.248 \pm 0.001 $. For
comparison, covariance of the measured signal and idler
photocounts was $0.238 \pm 0.001$ \cite{Brida2010}.

On the other hand and analyzing the data along the developed
method, we obtain the graph of minimum declinations $ {\cal D} $
depending on QDEs $ \eta_s $ and $ \eta_i $ as shown in
Fig.~\ref{fig1}.
The plotted minimum values of declinations $ {\cal D} $ are found
after scanning over the mean photon-pair number $ \langle n_p
\rangle $. The minimum value of declinations $ {\cal D} $ has
occured for QDEs $ \eta_s = 0.243 \pm 0.001 $ and $ \eta_i = 0.235
\pm 0.001 $. Whereas efficiency $ \eta_s $ gives directly QDE of
the camera, efficiency $ \eta_i $ encompasses also reflectivity of
the mirror placed in the path of the idler beam in perfect
agreement with its independently measured value. The comparison
with the values written above shows that partitioning the noisy
parts of the probe field results in lower values of QDEs. The
method has also provided the complete characteristics of the probe
field with 1\% relative error: $ b_p=0.058 $, $ M_p = 170 $, $
b_s= 33.2 $, $ M_s =0.0007 $, $ b_i= 10.6 $, and $ M_i= 0.0101 $.
Thus, the probe field has contained on average 9.91 pairs, 0.02
signal noise photons and 0.11 idler noise photons. Despite the
fact that the probe field has been nearly-exclusively composed of
photon pairs (for the JPND, see Fig.~\ref{fig1}), the effect of
noise on the values of QDEs cannot be neglected. It holds that the
larger the mean photon-pair number $ \langle n_p \rangle $ the
closer the values of QDEs obtained with the standard and the
developed methods.

In conclusion, we have developed a method for precise
determination of absolute detector efficiency of any photon-number
resolving detector. The method allows to partition noise from the
probe predominantly paired field thus providing, in principle, the
precision in detector calibration limited only by the number of
measurement repetitions. The improved precision and the
possibility to use noisier probe fields make the method superior
above other methods developed so far.

Support by projects P205/12/0382 of GA \v{C}R and COST OC 09026
and CZ.1.05/2.1.00/03.0058 of the Ministry of Education of the Czech Republic is
acknowledged. J.P.Jr. thanks J. Pe\v{r}ina for discussions.


\end{document}